\documentclass[graybox]{svmult}

\usepackage{type1cm}        
\usepackage{makeidx}         
\usepackage{graphicx}        
\usepackage{multicol}        
\usepackage[bottom]{footmisc}

\usepackage{newtxtext}       %
\usepackage[varvw]{newtxmath}       

\makeindex             

\begin{document}

\title*{Origin of cosmological density perturbations \\ 
from quantum fluctuations of vacuum \\ 
in General Relativity}
\titlerunning{Origin of cosmological density perturbations...} 
\author{V.N. Lukash\orcidID{0009-0009-3441-2576} and \\ 
E.V. Mikheeva\orcidID{0000-0002-3342-3259}}
\institute{V.N. Lukash \at ASC FIAN, Profsoyuznaya 84/32 117997 Moscow Russia \email{lukash@asc.rssi.ru}
\and E.V. Mikheeva \at ASC FIAN, Profsoyuznaya 84/32 117997 Moscow Russia \email{helen@asc.rssi.ru}}
%
%
\maketitle

\begin{svgraybox}

This book is dedicated to the memory of Alexei Alexandrovich Starobinsky, whose ideas and papers formed important part of modern cosmology. The origin of the Universe was one of problems he was interested in.

For many years Alexei adhered to the idea that the initial state of the Universe corresponds to the de Sitter world with its inherent high degree of symmetry \cite{star1}. The symmetry breaking transforms the world to the Friedmannian one which is observable now on large cosmological scales. Alexei associated the transition from the de Sitter model to the Friedman model with the extension of GR by introducing the $R^2$ term in the gravitational action (see \cite{star2} and references therein). 

In this paper we consider another concept of the evolving Universe, based on the idea of many fields in the vacuum state, where their initial expectation values are zero, but the total energy density is larger than zero. We call this state the polarized vacuum in GR. The exit from this state and ``rolling down'' successively in directions determined by different fields is responsible for the evolution of the Universe and can be tested through investigation of the  power spectra of cosmological perturbations. 

\end{svgraybox}

\abstract*{After 45 years since the discovery of quantum-gravitational birth of the cosmological density perturbations we can try to answer the main question of cosmology what is the origin of the Universe. This has become possible because the observational data are precise enough to build a physical model of the cascade relaxation of the gravitating vacuum as a generator of the evolving Universe. We present a model of the early Universe, which is inspired by the observational data in the spatial wavenumbers $k\in(2\cdot10^{-4}, 10)$ Mpc$^{-1}$ and does not require a solution of the Friedman equations. Based on observations we present a solution in the form of a vacuum attractor of General Relativity, which provides an additional power in the form of a ``bump'' and a blue spectrum of perturbations at $k>10\,$Mpc$^{-1}$.  
Extending the power spectrum to the scale of hundreds of kiloparsecs and less can solve the problems of the observed cosmology --- the appearance of the early star formation and supermassive black holes at $z>10$, the birth of primordial black holes, $\Lambda$CDM model, and others.}

\abstract{After 45 years since the discovery of quantum-gravitational birth of the cosmological density perturbations we can try to answer the main question of cosmology what is the origin of the Universe. This has become possible because the observational data are precise enough to build a physical model of the cascade relaxation of the gravitating vacuum as a generator of the evolving Universe. We present a model of the early Universe, which is inspired by the observational data in the spatial wavenumbers $k\in(2\cdot10^{-4}, 10)$ Mpc$^{-1}$ and does not require a solution of the Friedman equations. Based on observations we present a solution in the form of a vacuum attractor of General Relativity, which provides an additional power in the form of a ``bump'' and a blue spectrum of perturbations at $k>10\,$Mpc$^{-1}$.  
Extending the power spectrum to the scale of hundreds of kiloparsecs and less can solve the problems of the observed cosmology --- the appearance of the early star formation and supermassive black holes at $z>10$, the birth of primordial black holes, $\Lambda$CDM model, and others.}

\section{Introduction}

The equation of state of the matter ($E+P \simeq0$), responsible for the accelerated expansion of the Universe, is close to both the vacuum ($E+P =0$) and the scalar field equation of state with a dominant potential term ($\vert E+P\vert\ll E$). Interpretating the nature of the $\Lambda$-term as the Einstein's classical cosmological constant has led to an impasse, since the matter density cannot be stable. Another way leads to the concept of many fields, which is related to the idea of a large number of degrees of freedom (fundamental fields of matter) responsible for the polarization of vacuum in the Universe \cite{wing1, wing2, XXX1, book2010, rubakov, kazakov}. Since the first observational data were insufficiently precise their interpretation allowed for a large ambiguity in field interaction potentials. Different models based on those data led to different observational consequences. After 2021 (see \cite{BICEP}), the improved data precision is sufficient to build an observationally motivated model of the early Universe without solving the Friedman equations and without information about field potential (see \cite{LM}).

We understand by vacuum its polarization in the external gravitational field with its subsequent relaxation in time serving as a generator of the evolving Universe. In our model the solution of the vacuum attractor (VA) is suggested by the observational data obtained in the recent years \cite{LM}. The polarization fields initially were in zero states ($\varphi^{(i)}\simeq0$). At some moment one of them (having the largest mass among others) appears as a dominating field and begins to evolve from zero to some non-zero value, ending its motion in a new vacuum state (with energy density value less than the previous one). This process could repeat itself with the help of other fields in other times. It is the ongoing gravitational process of cascade relaxation of vacuum that creates all observable cosmology. The conditions for the appearance of the supermassive black holes (SMBH) \cite{SMBH} and early galaxies before the formation of the large-scale structure of the Universe \cite{JWST} are fulfilled when a non-power-law spectrum of cosmological density perturbations is considered. 
As high peaks (bumps) appear in the power spectra of density perturbations, primordial black holes (PBHs) can be borne in the Universe, as indicated by the LIGO data. The spectrum at small scales ($k > 10$ Mpc$^{-1}$) can be reconstructed if the observations of the J. Webb space telescope (JWST) are confirmed (see \cite{tkachev24a, eroshenko24, tkachev24b} and references therein). 

\section{The cascade relaxation of vacuum}

In GR the Lagrangian density has two terms: the first one consists of the metric tensor and its derivatives, $\frac{R}{16\pi G}$ (where $R = R_\mu^\mu$, $R_{\mu\nu}$ is the Ricci tensor, $G = const$), and the second one includes the fundamental degrees of freedom of the matter fields and their derivatives, 
\begin{equation}
\mathcal L=\mathcal L\left(w^{\left(1\right)},w^{\left(2\right)},..., \varphi^{\left(1\right)},\varphi^{\left(2\right)},...\right),
\label{eq1}
\end{equation}
where $w^{\left(1\right)2} = \varphi^{\left(1\right)}_{,\mu}\varphi^{\left(1\right),\mu}$, etc.\footnote{This includes all fields of all spins. To simplify the notation we do not use the space-time indices in the matter fields and their kinetic terms.}. The gravitational field is described by the metric tensor $g^{\mu\nu}$ and is included in the function (\ref{eq1}), but its derivatives are contained only in the scalar $R$. The separation of this kinetic scalar is the isolation of the gravitational field, which means the appearance of GR. A further decomposition of the Lagrangian with separation of different kinetic terms of the fields occurs evolutionarily, when energy decreases.

The energy density includes entropy, particles and vacuum polarization. The temperature and particle density decrease during the gravitational expansion of the Universe, resulting in 
domination of the vacuum polarization. The decrease of vacuum density occurs through a sequence of relaxation epochs, where one field dominates while the others remain gravitationally frozen.

 At the first stage the scalar field $\varphi$ dominates with simultaneous separation of its kinetic term from other fields:
\begin{equation}
\mathcal L\rightarrow \mathcal L\left(w,\varphi,...\right)\rightarrow\frac{w^2}{2}-V\left(\varphi,...\right),
\label{perehod}
\end{equation}
where $w^2 = \varphi_{,\mu}\varphi^{,\mu}$ is the kinetic scalar of field $\varphi = \varphi^{(1)}$, $V=V(\varphi,...)$ is a potential of all fields. In the initial (symmetric or basic) state all fields are zero. The first field starting the motion from zero dominates the potential (has the maximal value of mass). The kinetic scalars of other fields are zero, since they do not evolve due to the equations of motion.

The cascade relaxation of vacuum (CRV) process is considered as a transition of fields in time from symmetric state $V_0=V(0,0,...)$ to energetically lower states -- firstly to $V_1=V(\varphi_1,0,...) < V_0$ with field $\varphi$ (where not otherwise specified, we assume $\varphi > 0$), then from $V_1$ to $V_2 = V(\varphi_1, \varphi_2,0, ...)<V_1$ with field $\varphi^{(2)}$, etc.:
\[
V_0=V\left(0,0,...\right) \stackrel{\varphi}{\longrightarrow} V_1\left(\varphi_1,0,...\right)
\stackrel{\varphi^{\left(2\right)}}{\longrightarrow}
V_2\left(\varphi_1,\varphi_2,0,...\right) \longrightarrow ...
\]
CRV means the sequential epochs of domination of fields, including the present dark energy epoch. All states of the potential are positive since they are larger than the cosmological 
$\Lambda$-term, and it is positive from observations: 
\begin{equation}
V_0 > V_1 > V_2> ... > \Lambda > 0.
\end{equation}

We assume that the state $V_0$ is a polarization of vacuum of all fields. It results in the process of accelerated expansion of the Universe along all three spatial directions, that brings any initial space-time to the Friedman symmetry. The observational model of the first stage of the CRV creates all power spectra of cosmological perturbations.

\section{The early Universe model}
At the first stage of CRV, the field $\varphi=\varphi(x^\mu)$ emerged from the initial state $V_0$ ($\vert\varphi\vert \ll \varphi_1$), being the gravitational source of the evolving spatially flat Friedman model with linear fluctuations of the metric, which are described by the density perturbation field $q=q(x^\mu)$ and gravitational waves $\tilde q_{ij}=\tilde q_{ij}(x^ \mu)$ of all coordinates $x^\mu=(t,x^i)$. The solution is written as a power series in the small quantities $q$ and $\tilde q_{ij}$:
\begin{equation}
\varphi=\varphi\left(N\right)+\alpha\Delta+O\left(\Delta^2\right),
\label{phiN}
\end{equation}
\[
ds^2=\left(1+2\Phi\right)dt^2-a^2\left(1-2\Phi\right)\left(\delta_{ij}+2\tilde q_{ij}\right)dx^idx^j,
\label{metrika}
\]
where $\Delta=q-\Phi$ is matter velocity potential, $\Phi=\frac{H}{a}\int{\gamma q a\,dt}$ --- gravitational potential, $\tilde q_i^i=\tilde q_{i, j}^j=0$; $N=N(t)\equiv\ln(a)$, $\alpha =\alpha(N)\equiv\varphi_{,N}$, $H\equiv\dot N$ and $\gamma\equiv-\frac{\dot H}{H^2}$ depends on time, dot means the time $t$ derivative, $X_{,x}=\frac{\dot X}{\dot x}$ (for functions depending on time), increase and decrease of space indices $i$ and $j$ are performed by unit tensor $\delta_{ij}=\text{diag}(1,1,1)$.  

The field $\varphi(N)$ is a source of a homogeneous vacuum background satisfying the Friedman equations. The solution (\ref{phiN}) arises spontaneously during the gravitational expansion of a space-time region from the initial homogeneous scale $\sim1/H_i$ with the subsequent accelerated expansion of the boundaries $\sim a/(a_i H_i)$, which constrains the function $\gamma$: $\ddot a > 0 \rightarrow \gamma < 1$.

All fluctuations of the backgound can be described by two gravitational fields $q$ and $\tilde q_{ij}$ --- the scalar ($S$) and tensor ($T$) modes of linear perturbations of the metric, respectively. These are test  massless fields of the Friedman model with actions obtained by direct expansion of the general action to second order in $q$ and $\tilde q_{ij}$ \cite{Lukash80a, Lukash80b, book2010}:
\begin{equation}
\delta^{\left(2\right)}S = \int\left(L+\tilde L\right)\sqrt{-g}d^4x,
\end{equation}
\[
L=\frac{\gamma q_{,\mu}q^{,\mu}}{8\pi G} = \frac{\alpha^2q_{,\mu}q^{,\mu}}{2},\quad
\tilde L = \frac{\tilde q_{ij,\mu}\tilde q^{ij,\mu}}{16\pi G} =\frac{m_P^2\tilde q_{ij,\mu}\tilde q^{ij,\mu}}{4}, 
\]
where $m_P=\frac{M_P}{2\sqrt{\pi}}$, $M_P=\frac{1}{\sqrt G}$. Let us
remark that the field $q$ is included in the velocity potential of matter and has a Newtonian limit, the field $\tilde q_{ij}$ has two polarizations, the speed of the $S$-mode tends to the speed of light according to the condition (\ref{perehod}), and the functions $\alpha$ and $\gamma$ are related as $\gamma=4\pi G\alpha^2$.
In addition, it is convenient to introduce the following notation for the derivatives of $H$ and establish useful connections between them in the future:
\begin{equation}
\gamma = \beta^2 = -\frac{H_{,N}}{H},\quad
\beta=\frac{\alpha}{m_P} = \phi_{,N} = -\frac{H_{,\phi}}{H},\quad
\varepsilon = \beta_{,\phi},
\label{g5}
\end{equation}
where $\phi=\frac{\varphi}{m_P}$ is a dimensionless field.

The expectations of the fields $q$ and $\tilde q_{ij}$ are equal to zero, but there are non-zero dispersions:
\begin{equation}
\langle q^2\rangle=\int_0^\infty{q_k^2\,\frac{dk}{k}},\quad
\langle\tilde q_{ij}\tilde q^{ij}\rangle=\int_0^\infty{\tilde q_{k}^2\,\frac{dk}{k}},
\label{g6}
\end{equation}
where the brackets $\langle...\rangle$ mean averaging over vacuum states of the fields, the functions $q_k$ and $\tilde q_{k}$ are spectra of cosmological perturbations, $k$ is the wavenumber. When the scale of perturbations is larger than the cosmological horizon ($k<aH$), the fields $q$ and $\tilde q_{ij}$ are frozen and the spectra do not depend on time. The slopes (indices) of the spectra, their dependence on the scale and the ratio of the power spectra are determined as follows:
\[
n_k=\frac{d\ln q_k}{d\ln k},\quad 
\tilde n_k=\frac{d\ln\tilde q_k}{d\ln k},\quad 
A_k=\frac{d\ln\vert n_k\vert}{d\ln k},\quad 
r_k=\frac{\tilde q^2_k}{q_k^2},
\]
where $r_k$ by definition, four times less than the corresponding definition in \cite{Plancknew11}.

The observational data \cite{Pl20, BICEP} give the following values in the range of wavenumbers in $k\in(2\cdot10^{-4}, 10)$ Mpc${}^{-1}$:
\begin{equation}
q_k\simeq10^{-5}\left(\frac{k_c}{k}\right)^{n_c},\quad
n_c = 0.0175\pm0.0025,\quad
r_c < 10^{-2},
\label{qk1}
\end{equation}
where the boundaries of this range correspond to the present event horizon and the scale of dwarf galaxies, respectively, $c$ refers to the ``central'' wavenumber $k_c=0.05$, the wavenumber $k$ is measured in units of Mpc$^{-1}$ for the Hubble constant 67 km\,sec${}^{-1}$\,Mpc${}^{-1}$.

The cosmological scales corresponded to the event horizons $k=aH$ at the epoch of generation of $S$ and $T$ modes of perturbations due to their  causal evolution. The background model of the early Universe was described at that time by the dominant field $\phi$, which determined the function $H=H(\phi)$ and its derivatives $\gamma$, $\varepsilon$, and $\varepsilon_{,N}$. 
Assuming the last three functions to be small ($\ll\!1$, which is confirmed by observations) and leaving only the main terms on $\gamma$, we obtain the following perturbation spectra, indices, and ratio of power spectra at the scale $k=He^{N}$ (see (\ref{prilq})): 
\begin{equation}
q_k=\frac{H}{2\pi\vert\alpha\vert},\quad
\tilde q_k=\frac{H}{\pi m_P},
\label{qk2}
\end{equation}
\begin{equation}
n_k = -\frac{\varepsilon + \gamma}{1 - \gamma}\simeq -\varepsilon - \gamma,\quad r_k=4\gamma\simeq-4\tilde n_k,
\label{nk2}
\end{equation}
where $(\ln k)_{,N}=1-\gamma\simeq 1$, the index $k$ is omitted for the model functions. A comparison of (\ref{qk2}) and (\ref{nk2}) with (\ref{qk1}) shows that the functions $\gamma$, $\varepsilon$, and $\varepsilon_{,N}$ were indeed small, and the value of $\gamma$ did not exceed the error bars of the observations: 
\begin{equation}
\gamma<0.0025,\quad
\beta<0.05,\quad
\varepsilon+\gamma=0.0175\pm0.0025,
\end{equation}
where functions are evaluated at $k_c$. 

The present observational data provide a small ratio of the power spectra ($r_k<0.01$) and a constant index of the density perturbation spectrum in a certain wavenumber range ($n_k\simeq-0.02$). This allows us to consider the functions $\gamma$ and $\varepsilon$ as independent from each other and relate them directly to the observational data. By assuming the function $H=H(\phi)$ to be 
smooth\footnote{To obtain the perturbation spectra, we need the three first derivatives of the $H$-function ($\gamma$, $\varepsilon$, and $\varepsilon_{,N}$) and do not need the higher ones (see (\ref{g16})). By recovering the initial part of the $H(\phi)$ function (from zero with $\phi>0$) from observational data, the derivatives $\beta$, $\varepsilon_{,\phi}$, and the higher terms are still within the observational error bars.}, 
we obtain the observational model of the early Universe without solving the Friedman equation and without information about the dominant field potential. 

\section{The observational model of vacuum relaxation}
Starting from values close to zero, the field $\phi$ moves towards higher values, which allows us to present the observational model of vacuum relaxation (OMVR) as a series in the variable 
$\xi\equiv n_0\phi^2\ll1$:     
\begin{equation}
H=H_0 \left(1 - \frac{\xi}{2} + O \left(\xi^2\right)\right),\quad
N = \ln \left(\frac{k}{H}\right) = \ln \left(\frac{k}{k_c}\right)-N_c,
\label{eq14}
\end{equation}
\[
\gamma=n_0\xi\left(1 + O \left(\xi\right)\right),\quad
\varepsilon=n_0\left(1 + O \left(\xi\right)\right),\quad
\varepsilon_{,N} = n_0 O \left(\gamma\right),
\]
\[
q_k=\frac{\mathrm H_0\left(1+O \left(\xi\right)\right)}{2\pi n_0\phi}\simeq\frac{\mathrm H_0}{2\pi\beta_c}\left(\frac{k_c}{k}\right)^{n_0}, \quad
r_k = 4 n_0^2 \phi^2 \left(1+O \left(\xi\right) \right),
\]
\[
n_k = - \epsilon \left(1 + \xi\left(1 + n_0\right) + O \left(\xi^2\right)\right)
= - n_0 \left(1 + O \left(\xi\right) \right),\quad
\tilde n_k \simeq - n_0^2\phi^2,
\]
\[
\phi \simeq \phi_c e^{n_0\left(N + N_c\right)}\simeq\phi_c \left(\frac{k}{k_c}\right)^{n_0},
\quad 
N_c \simeq \ln\left(\frac{H_0}{k_c}\right),\quad
A_k = O \left(\gamma\right),
\]
where $H_0$ and $n_0$ are constants, $\mathrm H_{(0)}\equiv\frac{H_{(0)}}{m_P}$, 
$\beta_c = n_0\phi_c$. 
This model provides a power-law red spectrum at scales $k\in(2\cdot10^{-4}, 10)$ and is built on the current observational data:
\begin{equation}
\mathrm H_0\simeq10^{-6}\phi_c,\quad 
N_c \simeq 120 + \ln\phi_c,\quad 
n_0 \simeq 0.017,
\label{eq10}
\end{equation}
\[
\gamma_c<0.0025,\quad \beta_c<0.05,\quad \phi_c< 3.
\]

The OMVR has no free parameters, contains two known constants ($H_0$ and $n_0$) and relates to the time period with $\varphi<0.8\,M_P$, $\alpha<0.014\,M_P$ and 
$\xi\simeq\frac{\gamma}{\varepsilon}<0.17$.  
It describes the beginning of the first stage of vacuum relaxation, where the field was close to zero and moved toward larger values (remaining less than the Planck mass), which ensures the constancy of the power spectrum index and the smallness of the $\gamma$-function in the wavenumber range $k = (10^{-4},10)$ Mpc$^{-1}$. 
Other field trajectories (e.g., models with the field moving from large values to small values) relate the spectrum index to the $\gamma$-value and are not compartible to the observational data.  
The OMVR has been obtained from the present data and does not require further verification. 
To extend the OMVR to small scales ($k>10$), new data and information on the potential are needed (the counter terms associated with the third constant are within the observational error bars).  

\section{Quantum-gravitational creation of $S$ and $T$ modes of metric perturbations in the Friedman Universe}

The decomposition of the action $S=\int(-\frac{m_P^2}{4}R+\mathcal L)\sqrt{-g}d^4x$ on linear perturbations $q$ and $\tilde q_{ij}$ leads to $S=S^{(0)}+\delta^{\left(1\right)}S+\delta^{\left(2\right)}S$, where $S^{(0)}=\int(-\frac32m_P^2H^2+\mathcal L^{(0)})a^3dtd\bf x$, $\delta^{\left(1\right)}S$ nullifies on the Friedman equations, $\delta^{\left(2\right)}S=\int(L+\tilde L)a^3dtd\bf x$, the Lagrangian density of $S$-mode depends on two functions 
$\alpha^2=m_P^2\gamma$ and $c_S^{-2}=\frac{w\mathcal L_{,w,w}}{\mathcal L_{,w}}$, $L=\frac12\alpha^2(c_S^{-2}\dot q^2-q_{,i}q^{,i})$ (see \cite{Lukash80a, Lukash80b, book2010}). If the condition (2) is satified, then $c_S=1$.

Quantum fields of 4-coordinates $x^\mu=(t,x^i)$, $q$ and $\tilde q_{ij}$, and 4-momenta $k_\mu=(\frac ka,k_i)$,  $q_{\bf k}$ and $q_{{\bf k}\xi}$, in Eucledian 3-spaces of Friedmannian model ${\bf x}=(x^i)$ and ${\bf k}=(k_i)$, respectively, are related by the Fourier's transformation:
\begin{equation}
q=\int\limits_{-\infty}^{\infty} q_{\bf k}e^{i{\bf k}{\bf x}}\frac{d\bf k}{\left(2\pi\right)^{3/2}},\quad 
q_{\bf k}=\nu_ka_{\bf k}+\nu^*_ka^{\dagger}_{-\bf k},
\label{g14}
\end{equation}
\[
\tilde q_{ij}=\sum\limits_\sigma\int\limits_{-\infty}^{\infty} p_{ij\sigma} q_{{\bf k}\sigma}e^{i{\bf k}{\bf x}}\frac{d\bf k}{\left(2\pi\right)^{3/2}},\quad
q_{{\bf k}\sigma} = \tilde \nu_ka_{{\bf k}\sigma}+\tilde \nu^*_k a^{\dagger}_{-{\bf k}\sigma},
\] 
where $p_{ij\sigma}=p_{ij\sigma}(\bf k)$ are constant normalized polarization tensors of ${\bf k}$-waves ($p^i_{i\sigma}=p_{i\sigma}^jk_j=0$, $p^*_{ij\sigma}p^{ij}_{\sigma^\prime}=\delta_{\sigma\sigma^\prime}=\text{diag}(1,1)$, $\sigma=\oplus,\otimes$). 
The commutation relations of canonically conjugated scalars
 $q$ and $p=\frac{\partial L}{\partial\dot q}=\alpha^2\dot q$ in the position space determine the commutators of constant operators of annihilation, $a_{\bf k}$, and creation, $a^\dagger_{\bf k}$, of particles (phonons) in {\bf k}-space:
\[
\left[q\left(t,\bf x\right)p\left(t,{\bf x}^\prime\right)\right] = qp - pq = 
i\frac{\delta\left({\bf x} - {\bf x}^\prime\right)}{a^3},\quad 
\left[a_{\bf k}a^\dagger_{{\bf k}^\prime}\right] = \delta \left({\bf k} - {\bf k}^\prime\right),
\]
where $\delta=\delta(\bf x)$ is Dirac function. The evolution of $q_{\bf k}$-oscillators is described by the classical normalized function $\nu_k=\bar\nu_k/(\alpha a)$, which depends on time $\eta=\int\frac{dt}{a}=\int\frac{dN}{aH}$ and wavenumber $k=\vert\bf k\vert$:
\[
\bar\nu_k^{\prime\prime} + \left(k^2 -U \right) \bar\nu_k = 0,\quad 
U=\frac{\left(\alpha a\right)^{\prime\prime}}{\alpha a},\quad
\bar\nu_k\bar\nu_k^{*\prime}- \bar\nu_k^* \bar\nu_k^\prime=i.
\]
Similar relations are valid for operators of annihilation ($a_{{\bf k}\sigma}$) and creation ($a^\dagger_{{\bf k}\sigma}$) of gravitons and evolutionary functions of $q_{{\bf k}\sigma}$-oscilators, $\tilde\nu_k=\sqrt2\bar{\tilde\nu}_k/(m_Pa)$: 
\[
[a_{{\bf k}\sigma}a^\dagger_{{\bf k}^\prime\sigma^\prime}]=\delta \left({\bf k} - {\bf k}^\prime\right)\delta_{\sigma\sigma^\prime},\quad 
[a_{{\bf k}\sigma}a_{{\bf k}^\prime\sigma^\prime}]=0,
\]
\[
\bar{\tilde\nu}_k^{\prime\prime} + \left(k^2 - \tilde U\right) \bar{\tilde\nu}_k =0,\quad
\tilde U=\frac{a^{\prime\prime}}{a},\quad
\bar{\tilde\nu}_k\bar{\tilde\nu}_k^{*\prime}- \bar{\tilde\nu}_k^* \bar{\tilde\nu}_k^\prime=i.
\]
The cosmological $S$ and $T$ modes of zero oscilations of the Friedmannian model are determined  by the vacuum $\vert 0 \rangle$-state of all linear fields:
\[ 
a_{{\bf k}\left(\sigma\right)}\vert 0\rangle
=0, \quad
\langle a_{{\bf k}}a^\dagger_{{\bf k}^\prime}\rangle=\delta\left({\bf k}-{\bf k}^\prime\right),
\quad
\langle a_{{\bf k}\sigma}a^\dagger_{{\bf k}^\prime\sigma^\prime}\rangle=\delta({\bf k}-{\bf k}^\prime)\delta_{\sigma\sigma^\prime}. 
\]

The bilinear forms (\ref{g6}) are as follows (see (\ref{g14})): 
\begin{equation}
q_k = \frac{k^{3/2}\vert\nu_k\vert}{\sqrt 2\pi},\quad
\tilde q_k = \frac{k^{3/2}\vert\tilde\nu_k\vert}{\pi}, \quad 
r_k = \frac{\tilde q_k^2}{q_k^2} = \frac{2\vert\tilde\nu_k\vert^2}{\vert\nu_k\vert^2}.
\label{g15}
\end{equation}

The parametric potentials of elementary oscillators are following:
\begin{equation}
U = a^2 H^2 \left(\left(2 - \gamma + \varepsilon\right) \left(1 + \varepsilon\right) + \varepsilon_{,N}\right),\quad
\tilde U = a^2 H^2 \left(2 - \gamma\right).
\label{g16}
\end{equation}
They coincide when the following inequalities are satisfied,
\[
\gamma\ll1,\; \vert\varepsilon\vert\ll1,\;\vert\varepsilon_{,N}\vert\ll1:\qquad 
U = \tilde U = \frac{2}{\eta^2},\quad \eta=-\frac{1}{aH},
\]
which solves the problem of cosmological perturbations in Fridman model:
\begin{equation}
\bar\nu_k=\bar{\tilde\nu}_k=\frac{e^{-ik\eta}}{\sqrt{2k}}\left(1+\frac{1}{ik\eta}\right),\quad r_k=4\gamma,
\label{prilq}
\end{equation}
\[
q_k=\frac{\sqrt{H^2+\frac{k^2}{a^2}}}{2\pi\vert\alpha\vert} \longrightarrow
\frac{\mathrm H}{2\pi\vert\beta\vert},\quad 
\tilde q_k=\frac{\sqrt{H^2+\frac{k^2}{a^2}}}{\pi m_P}\longrightarrow
\frac{\mathrm H}{\pi}.
\]
The last limit means time-frozen power spectra in wavelengths $k\ll aH$, where the functions $H=m_P\mathrm H$ and $\alpha=m_P\beta$ are calculated at the moment of time $\eta=-1/k$.

\section{A vacuum attractor}

Since the fields evolve from zero values, the potential can be decomposed over the powers of the fields. Let us assume that  the first three terms are essential and the rest is small (see \cite{LM}): 
\begin{equation}
V=V_0-\frac{m^2\varphi^2}{2}+\frac{\lambda\varphi^4}{4}+...\longrightarrow V_1+m^2\left(\frac{\varphi^2-\varphi_1^2}{2\varphi_1}\right)^2,
\label{3const}
\end{equation}
where $V_{0,1}$ and $\varphi_1=\frac{m}{\sqrt\lambda}$ are positive constants (the coefficients of the series are determined by all fields). Owing to symmetry of $V$ relative to the sign of $\varphi$, the field evolves from the symmetric state $V_0=V(0)$ to a positive (or negative) $\varphi$ with a lower-energy state $V_1=V(\pm\varphi_1)$ (see Fig.~1):
\begin{equation}
V_0=V_1+\frac{m^4}{4\lambda}>V_1>0.
\end{equation}
%

\begin{figure}[b]
\sidecaption
\includegraphics[width=0.64\textwidth]{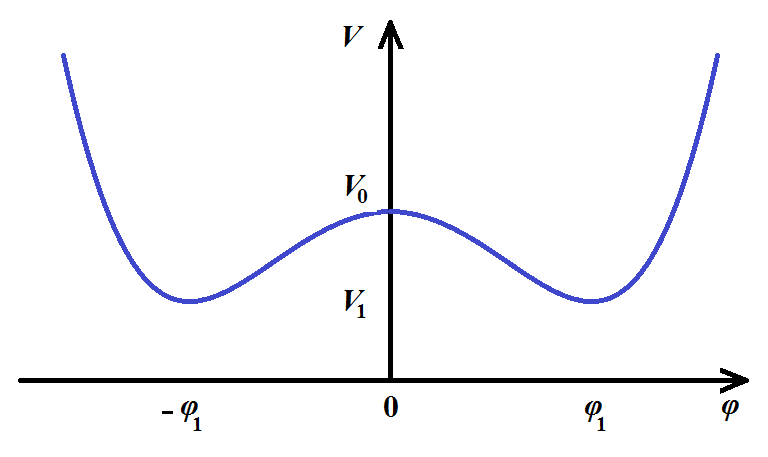}
\caption{The potential $V$ as function of $\varphi$. A minimal value $V_1$ corresponds to $\varphi=\pm\varphi_1$}
\label{fig1}
\end{figure}

During the rolling down of the field along the potential the height of the vacuum step $V_{01}= V_0-V_1$ depends on two fundamental constants, $V_{01}=\frac{m^4}{4\lambda}=\frac{m^2\varphi_1^2}{4}$. Since there were three constants in (\ref{3const}), we take as the third (independent) one the value of the residual vacuum in units of the step height, $v_1=\frac{V_1}{V_{01}}$, or the amplitude of the initial vacuum, $\mathrm v_1=\frac{V_1}{V_0}\in(0,1)$:
\begin{equation}
v=v_1 + y^2,\quad 
\mathrm v=\mathrm v_1 + \frac{y^2}{v_0} = 1 - \frac{\mathrm n_0\left(1+y\right)\phi^2}{2},
\end{equation}
where $v=\frac{V}{V_{01}}$, 
$\mathrm v=\frac{v}{v_0}$, 
$y=1-x^2$ and 
$x=\frac{\varphi}{\varphi_1}=\frac{\phi}{\phi_1}$ are dimensionless, $v_0=1+v_1=(1-\mathrm v_1)^{-1}$, $\mathrm n_0=\frac{m_P^2m^2}{2\mathrm V_0}$, $\phi_1=\frac{\varphi_1}{m_P}=\sqrt{\frac{2}{v_0\mathrm n_0}}$. At $v_1> 1$, $\mathrm v \simeq 1$ in the interval $x\in(0,1)$. 
When $v_1<1$, the function $\mathrm v$ varies from $\mathrm v\simeq 1$ at $y\in (1,\frac{1}{\sqrt2})$ to $\mathrm v\simeq\mathrm v_1$ at $\vert y\vert<\sqrt{\frac{\mathrm v_1}{1+\mathrm v_1}}$, decreasing at $y^2\in(\frac12,\frac{\mathrm v_1}{1+\mathrm v_1})$.

Equations of motion can be derived by varying the action over $g^{\mu\nu}$ and $\varphi$:
\[
\phi_{,\mu}\phi^{,\nu}+\left(\mathrm V-\frac{\mathrm w^2}{2}\right)\delta_\mu^\nu=\frac12G_\mu^\nu,\quad 
x^{;\mu}_{;\mu}+m^2x\left(x^2 - 1\right) = 0,
\]
where $\mathrm V_{(0,1)} = \frac{V_{(0,1)}}{m_P^2}$, $\mathrm w^2=\phi_{,\mu}\phi^{,\mu}$,  $G_\mu^\nu=R_\mu^\nu-\frac 12R\delta_\mu^\nu$, the equation for $\varphi$ is included in gravity equations due to the Bianchi identities ($G_{\mu;\nu}^\nu=0$). 

For the case of homogeneous field the Friedman equations are as follows:
\begin{equation}
H^2=\frac{\dot\phi^2+2\mathrm V}{3}=\frac{2\mathrm V}{3-\gamma},\quad
\dot H=-\dot\phi^2,
\end{equation}
resulting in the equation of motion of $\phi=\phi_1x$ and/or $\dot\phi=H\beta$:
\begin{equation}
\ddot x+3H\dot x+m^2x\left(x^2 - 1\right) = 0,\quad
\frac{\beta\beta_{,\phi}}{3 - \beta^2} + \beta = \frac{\mathrm n_0y\phi}{\mathrm v},
\label{eqmain}
\end{equation}
\[
H = H_0\sqrt{\frac{\mathrm v}{1 - \frac{\gamma}{3}}} = \frac{2\mathrm m}{3}\sqrt{\frac{\bf v}{1 -\frac{\gamma}{3}}}, \quad
\gamma = \frac{6\dot x^2}{2\dot x^2+m^2v}
\in\left(0,3\right)\!,
\]
where $H_0 = \sqrt{\frac{2\mathrm V_0}{3}} = \frac{m}{\sqrt{3\mathrm n_0}}$, $\mathrm m=\sqrt2m$, ${ \bf v}_{(1)}=\frac{3\mathrm  v_{(1)}}{8\mathrm n_0}$,
$H_1=\sqrt{\frac{2\mathrm V_1}{3}}=H_0\sqrt{\mathrm v_1} = \frac{2\mathrm m}{3}\sqrt{{\bf v}_1}$. 
The equation for $x$ can be rewriten in terms of $\mathrm x=1-x$, which is useful to describe the solutions of the residual vacuum:
\begin{equation}
\ddot{\mathrm x}+3H\dot{\mathrm x}+m^2\mathrm x\left(1-\mathrm x\right)\left(2-\mathrm x\right)=0.
\label{eqx}
\end{equation}
Let's represent this equation in terms of $X=xe^{\frac32N}$ and $\mathrm X=\mathrm xe^{\frac32N}$:
\begin{equation}
\ddot X - \mathrm m^2{\bf v}_+ X=0, \quad
\ddot{\mathrm X}+\mathrm m^2{\bf v}_-\mathrm X=0,
\label{eqX}
\end{equation}
where 
${\bf v}_+ = \frac y2+\bar{\bf v}$, 
${\bf v}_- = \frac{x(1+x)}{2}-\bar{\bf v}$, 
$\bar{\bf v} = {\bf v}(\frac{3-2\gamma}{3-\gamma})$. 
As we change from $t$ to $mt$, the mass parameter drops out of the equations ($m$ determines time/energy), and the evolution depends on the two constants $\mathrm n_0$ and $\mathrm v_1$.

The solutions of the equation for $\phi$ are trajectories on the phase plane $(\phi,\dot\phi)$, among which there are three special points (poles): the central $\phi=\dot\phi=0$ and side $\phi\pm\phi_1=\dot\phi=0$, where $\gamma=\beta=\varepsilon_{,N}=0$, $H=H_{0,1}$ and $\varepsilon=\varepsilon_{0,1}$, respectively. At the center pole, two parameters $H_0$ and $\mathrm n_0$ are given, while the third $\mathrm v_1$ is not included in the derivatives (\ref{g5}). All three parameters $H_1$, $\mathrm v_1$ and $\mathrm n_1\!=\frac{2\mathrm n_0}{\mathrm v_1}=\frac{3}{4{\bf v}_1}$ are necessary at the side poles. The constants $\varepsilon_{0,1}$ are related to the vacuum constants $\mathrm n_{0,1}$ by means of two independent binomials (see eqs.~(6) and (\ref{eqmain})):
\begin{equation}
\varepsilon_0^2+3\varepsilon_0-3\mathrm n_0=0,\quad \varepsilon_1^2+3\varepsilon_1+3\mathrm n_1=0.
\end{equation}
If $\varepsilon_0>0$ the equation has the only solution
\[
\varepsilon_{0+}\equiv n_0=\frac{3\mathrm n_0}{\vert\varepsilon_{0-}\vert}=\frac{2\mathrm n_0}{1+\sqrt{1+\frac{4\mathrm n_0}{3}}},
\]
where $\varepsilon_{0\pm}=\frac32(-1\pm\sqrt{1+\frac{4\mathrm n_0}{3}})$. The second root of equation, $\varepsilon_{0-}$, is negative, since the binomial is invariant under the transformation $\varepsilon_0\rightarrow-3-\varepsilon_0$. At ${\bf v}_1>1$ $(\mathrm n_1<\frac34)$ both roots are negative, $\varepsilon_{1\pm}=\frac32(-1\pm\sqrt{1-{\bf v}_1^{-1}})$, what is needed to build the solution of the vacuum 
attractor\footnote{Two trajectories 
enter/exit to/from the central pole anong the axes $\dot\phi=-(3+n_0)H_0\phi$ and $\dot\phi=n_0H_0\phi$, respectively. Other trajectories, approaching the point $\dot x=x=0$, turn without entering the pole. All trajectories, except for the central point, enter the side (stable) poles.}.

The solution of the equations can be found in integral form:
\begin{equation}
\phi=C\exp\left(\int ndN\right),\quad 
\beta=n\phi,\quad
\varepsilon=\left(nx\right)_{,x},
\end{equation}
where $C=const$. Equation for $n=n(x)$ is as follows
\[
n_{,N}=nn_{,x}x=\left(3-\gamma\right)\!\left(\mathrm n-n\right)-n^2 =
\left(n_ + -n\right) \left(n-n_-\right),
\]
where $\mathrm n=\frac{\mathrm n_0 y}{\mathrm v}$, 
$n_{\pm}=\frac{3-\gamma}{2}\left(-1\pm\sqrt{1+\frac{4\mathrm n}{3-\gamma}}\right)$.
The functions $n$ and $\mathrm n$ are related algebraically via the function $\epsilon$, 
\begin{equation}
n=\frac{\mathrm n}{1+\frac{\epsilon}{3}},\qquad
\epsilon=\frac{\varepsilon}{1-\frac{\gamma}{3}},
\end{equation}
which satisfies the equation ($\epsilon_{,N}=n\epsilon_{,x}x$):
\[
\epsilon_{,N} =
3{\bf n} + \left(\gamma - 3\right) \left(\epsilon + \frac{\epsilon^2}{3}\right) =
\left(1 - \frac{\gamma}{3}\right) \left(\epsilon_+ - \epsilon\right) \left(\epsilon - \epsilon_- \right),
\]
where ${\bf n} = (\mathrm nx)_{,x} = \frac{\mathrm n_0{\bf y}}{\mathrm v}$, 
$\epsilon_{\pm}=\frac{\varepsilon_\pm}{1-\frac{\gamma}{3}}=\frac 32(-1\pm\sqrt{1+\frac{4{\bf n}}{3-\gamma}})$, 
${\bf y}=\mathrm v(\frac{xy}{\mathrm v})_{,x} = 1 + \frac{x^2(\mathrm v-4\mathrm v_1)}{\mathrm v}$. 

In the interval $x\in(0,1)$ the function ${\bf y}$ varies from 1 to -2, crossing 0  
(${\bf y}_*=0$) at $x_*^2=\frac{ly_*}{2}=\frac{l}{l+2} \in (\frac13,1)$:
\[
{\bf y}=\frac{\hat y\left(c_0 c_1 + \left(l-1\right)\tilde y - \tilde y^2\right)}{\left(l + 2\right)\left(c_1 + 2\tilde y + \tilde y^2\right)} = c\hat y + O \left(\hat y^2 \right),
\]
where $\tilde y=\frac{y-y_*}{y_*}$, $\hat y = 2\tilde y$, $c = \frac{c_0}{l + 2}\in (0.7,1)$. The constants are determined by the parameter $l>1$,
\[
c_0=l+1+l^{-1},\quad 
c_1=\frac{2l}{l-1},\quad 
\mathrm v_1 = \frac{\mathrm v_*\left(l+1\right)}{2l} = \frac{4\left(l+1\right)}{l\ell},
\]
\[
v_1 = \frac{y_*^2 \left(l + 1\right)}{l-1}, \quad \ell = l^2 + 3l + 4 = \left(l_+ + 2\right)\left(l_- + 2\right),\quad l_\pm = l\pm\sqrt{l}.
\]

The general solution has the following approximation:
\begin{equation}
n=\frac{n_0y}{u\mathrm v},\quad 
\beta = \frac{\beta_1xy}{u\mathrm v},\quad 
\epsilon=3\left(u - 1\right) + n_0 u = \frac{n_0 \left({\bf y} - \frac{u_{,x}xy}{u}\right)}{u\mathrm v\left(1 - \frac{\gamma}{3}\right)},
\end{equation}
where $u = \frac{3 + \epsilon}{3 + n_0}$, $\beta_1 = \sqrt{\gamma_1} = n_0 \phi_1 = \sqrt{\frac{2\bf n_0}{v_0}}$, ${\bf n}_0 = \frac{3n_0}{3+n_0}$.

A vacuum attractor (VA) is a partial solution of the Friedman equations consisting of a sequence of eras: the motion goes from the quantum boundary to the central pole, approaches it along the axis $\dot\phi = -(3+n_0)H_0\phi$ and exits along $\dot\phi=n_0H_0\phi$, where $n=\epsilon= \varepsilon= n_\pm= \epsilon_\pm= (n_0,-3-n_0)$ near $\phi=\dot\phi=0$. 

The exit of VA from the center to the region $\phi>0$ corresponds to the OMVR (\ref{eq14}) with $\xi=\frac{6x^2}{v_0(3+n_0)}$, where the constants $H_0$ and $n_0$ are identical to those described in (\ref{eq10}).   
Using the smallness of $n_0$, we represent the functions of $x\in(0,1)$ as a series over 
$\zeta = x\bar\zeta = \frac{{\bf n}_0x^2}{\mathrm v}$\,\footnote{The constants $n_0$, 
$\mathrm n_0$, and ${\bf n}_0$ are arranged from observations, $\sqrt{\mathrm n_0{\bf n}_0}=n_0$, 
$\sqrt{\mathrm n_0/{\bf n}_0}=1+\frac{n_0}{3}$. The squares of the variables $\zeta$ and $\bar\zeta$ are as follows: $\zeta^2<\zeta\bar\zeta< \bar\zeta^2<\zeta<\bar\zeta<1$.} 
with $\gamma<1$:
\begin{equation}
u=1+\mathrm u\zeta+\frac{{\bf u}\bar\zeta^2}{3}+O\left(\frac{n_0^3x^2}{\mathrm v^3}\right),\quad\gamma=\frac{2\zeta y^2}{u^2\left(y^2+v_1\right)},
\label{eq35}
\end{equation}
\[
\epsilon=\frac{n_0}{\mathrm v}\left({\bf y}+{\bf u}\zeta + O\left(\bar\zeta^2\right)\right),\quad
\phi\simeq\bar{\phi_0}y^{\mathrm v_1 /2}\left(\frac{k\,e^{\phi^2/2}}{k_0\sqrt{\mathrm v}}\right)^{n_0},
\]
\[
N + N_0 \simeq \frac{1}{n_0}\left(\ln \vert x\vert-\frac{\mathrm v_1}{2}\ln\vert y\vert + \frac{2y - 1 +\ln2}{2v_0}\right) = \ln \left(\frac{k}{k_0\sqrt{\mathrm v}}\right),
\]
where functions $\mathrm u$ and $\bf u$ are 
constrained\,\footnote{The formulas for these functions are as follows:
\[
u =1+ \frac{\epsilon-n_0}{3+n_0}, \quad
\mathrm u=\frac{{\bf y}-\mathrm v}{3x^2}=1-\frac{2\mathrm v_1\left(\mathrm v+2\right)}{3\mathrm v}-\frac{\mathrm n_0\phi^2}{6},
\]
\[
{\bf u}={\bf y}\bar{\mathrm u}+\frac{2{\bf y}_{,y}y}{3},\quad
\bar{\mathrm u}=\frac13\left(-5+\frac{2\mathrm v_1\left(\mathrm v+3\right)}{\mathrm v}+\frac{\mathrm n_0\phi^2}{2}\right).
\]
}, 
$\vert\mathrm u\vert<1$, $\vert{\bf u}\vert<1$, the condition $k=\frac{k_0\sqrt{1+3\mathrm v_1}}{2}$ ($N=-N_0$) is valid at the point $y =x^2 = \frac12$ ($\phi=\phi_0$), the constants are related: $\bar{\phi_0}=\phi_1(\sqrt{2e})^{\mathrm v_1-1}$, $N_0 = \ln(\frac{H_0}{k_0})$, $\phi_0 = \frac{\phi_1}{\sqrt2} = \frac{1}{\sqrt{v_0 n_0}} \simeq 7\sqrt{1 - \mathrm v_1}$, $\phi_1 \simeq 10\sqrt{1 - \mathrm v_1}$. The series of VA are valid for $x < \frac 89$, $y>\frac29$. For smaller $y$, when the solution of the VA is close to the pole $x - 1= \dot x = 0$, we need to consider eq. ~(\ref{eqx}).

The potential of all fields responsible for the vacuum polarization is obtained as a series over the fields with their values close to zero. At the first stage of the CRV, the first three terms of the series of the dominant field are taken without increasing the number of constants of the potential (the higher terms were assumed 
negligible)\footnote{The subject of our model is not the way of reduction of the cosmological constant by means of introduction of a ``specific'' potential of interactions. Thus, the potential of the \cite{Abbott85} model contains more constants and higher derivatives of the field (in the form of the field cosine), and the movement from large to small fields does not correspond to our approach. By vacuum we understand its polarization in the external gravitational field with the subsequent relaxation of vacuum in time as a generator of the evolving Universe. }. 
There are three coefficients of the three terms of the  series (\ref{3const}) which give three constants of the VA: $H_0$, $n_0$ and $\mathrm v_1$. The first two are determined from observations, and the third is free -- it does not affect the onset of VA and appears later in time, at $k>10$, when the solution changes. The functions $n$ and $\varepsilon$ are close initially ($\phi<3$) and then, approaching $\phi_1$, they diverge according to the ratio of the parameters $\mathrm v_1$ and $\mathrm v_{cr}=\frac{8\mathrm n_0}{3}\simeq0. 05$, ${\bf v}_1=\frac{\mathrm v_1}{\mathrm v_{cr}}\simeq20\mathrm v_1$ (see Fig.~2).

\begin{figure}
\sidecaption
\includegraphics[width=0.64\textwidth]{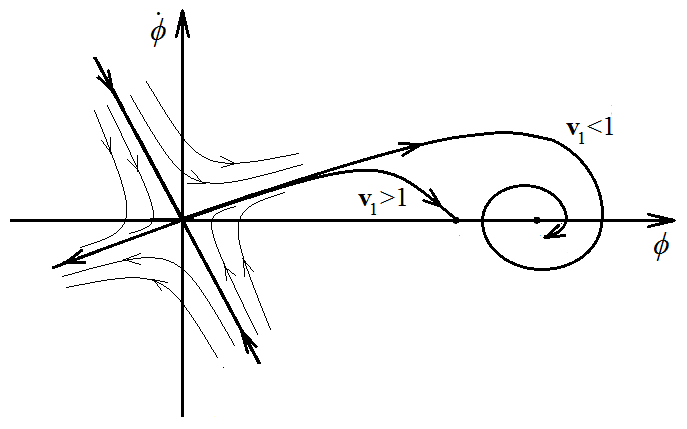}
\caption{The trajectories of the solution of eq. ~(\ref{eqmain}) on the phase plane 
($\phi$, $\dot\phi$). Bold lines correspond to the vacuum attractor, thin lines 
to the general solution, and arrows indicate the time direction (see \cite{LM})}
\label{fig2}
\end{figure}

\section{The case of a small residual vacuum $({\bf v}_1<1)$}
The function $\gamma$ increases monotonically in time, approaching $\gamma\sim1$ at $t\simeq\mathrm m^{-1}$ ($x\simeq0.87$, $y\simeq\sqrt{\mathrm v_{cr}}\simeq0.2$, $H\simeq\frac{2\mathrm m}{3}$). If $t>\mathrm m^{-1}$ ($y<0.2$), the VA trajectory circles the pole clockwise, crossing the $x=1$ axis at time $t_1$, and the $\dot x=0$ axis at $t_b>t_1$, moving from the top to the bottom of the phase plane. Then the field continues to oscillate around the point $y=\dot y=0$, approaching it: $y\simeq0. 2\,\frac{\cos(\mathrm mt)}{\mathrm mt}$ ($H\simeq\frac{2}{3t}$) at $t\in(\mathrm m^{-1},t_{\mathrm v})$, and $y \sim e^{-t/t_{\mathrm v}}\cos(\mathrm m_{\mathrm v}t)$ ($H\simeq\frac{2}{3t_{\mathrm v}}$) when $t>t_{\mathrm v}$, where $t_{\mathrm v}=\frac{1}{\mathrm m\sqrt{{\bf v}_1}}$, $\mathrm m_{\mathrm v}=\mathrm m\sqrt{1-{\bf v}_1}$, we omit the $\sim1$ phases of the cosines. Here the first stage of the cascade terminates and the problem is reduced to the previous one - the second stage of the CRV with a new vacuum $V=V_1$ and a new dominating field appears.

\section{The case of higher residual vacuum $({\bf v}_1>1)$}

Here $\mathrm v_1=\frac{4(l+1)}{l\ell}\in(\frac{1}{20},1)$, $l\in (1,8)$, $v_1=\frac{4(l+1)}{(l+2)^2(l-1)}>\frac{1}{20}$, $\sqrt{\mathrm v_1}\in(\frac29,1)$. 
The function $\gamma$ increases with the field and reaches the maximum value at $\gamma_r = \beta_r^2 \simeq \frac{n_0(l - 1)(l + 2)}{9} \simeq \frac{0,007}{v_1}$ at $x_r \simeq \sqrt{\frac{l}{l + 2}} \in (0.6,0.9)$ and goes back to zero at $x\rightarrow1$, remaining small at $x\in(0,1)$. At $x\in(0,\frac89)$, $y\in(1,\frac29)$ there is $\beta\simeq\frac{2xy}{\phi_1v}\le\beta_r\simeq\frac{0.08}{\sqrt{v_1}}\in(0, 0.4)$, $\dot x\simeq\frac{\mathrm m xy}{\sqrt{3v}}$. In the region $\mathrm x\in(\frac19,\frac12\sqrt{\mathrm v_1})$ there is $\beta\simeq\frac{4\mathrm x}{\phi_1v_1}$ and $\dot x\simeq\frac{2\mathrm m\mathrm x}{\sqrt{3v_1}}$. For $\vert\mathrm x\vert<\frac12\sqrt{\mathrm v_1}$ the solution (\ref{eqx}) is as follows:
\begin{equation}
\mathrm x = \mathrm x_{\times} \left(\mathrm c_ + e^{-\bar t_+} + \mathrm c_- e^{- \bar t_- }\right)=\mathrm x_{\times}e^{-\bar t}\left(\textrm{ch}\left(\omega\bar t\right)-
\frac{\mathrm c \cdot \textrm{sh} \left(\omega\bar t\right)}{\omega}\right),
\label{eq36}
\end{equation}
\[
\frac{\dot x}{\mathrm m\sqrt{{\bf v}_1}\mathrm x}=1+\omega+\frac{2\omega}{{\bf c}e^{-2\omega\bar t}-1}=
1+\mathrm c+
\frac{\left(\mathrm c^2 - \omega^2\right)\textrm{th}\left(\omega\bar t\right)}{\omega-\mathrm c \cdot \textrm{th}\left(\omega\bar t\right)},
\]
where $\bar t=\mathrm m\sqrt{{\bf v}_1}(t-t_{\times})\geq0$ and $\bar t_{\pm}=(1\pm\omega)\bar t$ are the variables, $\mathrm x_\times$, $\omega=\sqrt{1-{\bf v}_1^{-1}}$, $\mathrm {c}_\pm=\frac{\omega\pm\mathrm c}{2\omega}$, ${\bf c}=\frac{\mathrm c+\omega}{\mathrm c-\omega}$ and $\mathrm c$ are constants.The quantities $\mathrm x_{\times} \in (\frac19, \frac12\sqrt{\mathrm v_1})$ and $\mathrm c\simeq\frac{2}{\sqrt{3v_1{\bf v}_1}} -1 \in (4.2,-1)$ can be derived from the cross-linking with VA at $\bar t=0$ ($\mathrm x=\mathrm x_{\times}$, $\dot{\mathrm x}\simeq\frac{2\mathrm m}{\sqrt{3v_1}}\mathrm x_{\times}$).

Let us consider two cases.

For $t > t_\times$ and $\mathrm c>\omega$ ($v_1<\frac17$, $\mathrm v_1<\frac18$, ${\bf v}_1 \in (1, 2.7)$, $\omega \in (0,0.8)$, $ l \in (4.5,8)$)\footnote{The condition $c^2=\omega^2$ leads to the equation $\frac{4}{3v_1}+1=\phi_1\simeq\frac{10.7}{\sqrt{v_0}}$, whence we obtain $v_1\simeq\frac17$, $\omega \simeq \frac 45$ (the second root of equation $v_1 \simeq 100$ is rejected).}, 
the trajectory of VA circles the pole $\mathrm x=\dot{\mathrm x}=0$ clockwise,  at $\bar t_1 = \frac{\ln{\bf c}}{2\omega}$ it crosses the axis $x=1$ and moves in the upper halfplane. At $\bar t_b$ the trajectory crosses the axis $\dot x=0$ moving from upper area to down area of the plane, and then at $\bar t > \bar t_2$ it goes to the pole from the right 
($\phi>\phi_1$) along the axis $\dot\phi = \frac{\mathrm m(\phi_1-\phi)}{\sqrt{{\bf v}_1} + \sqrt{{\bf v}_1-1}}$, 
where $\bar t_{1,2} = \bar t_b\pm\Delta_b$, $\Delta_b=\frac{\ln (\frac{1+\omega}{1-\omega})}{2\omega} = \frac{\ln(\sqrt{{\bf v}_1}+\sqrt{{\bf v}_1-1})}{\omega}$. In this interval the function $n$ approaches zero,  and $\varepsilon$ is finite: $\varepsilon\simeq(\ln\vert\dot x\vert)_{,N}$. It is constant at $\bar t < \bar t_1$ ($\varepsilon\simeq-n_1$) and $\bar t>\bar t_2$ ($\varepsilon \simeq-n_2$, where 
$n_2=\frac{2n_1}{1+\omega}$)\,\footnote{From (\ref{eq36}) one can derive 
\[
\mathrm x\propto\left({\bf c} e^{-\tilde t} - 1\right) e^{-\bar t_-},
\quad \dot x \propto \left(e^{\tilde t_b-\tilde t} - 1\right) e^{-\bar t_-},
\quad \ddot x \propto \left(1 - e^{\tilde t_2-\tilde t}\right)\!e^{-\bar t_-},
\]
where $\tilde t = 2 \omega \bar t$, $\tilde t_b=\ln{\bf c}+2\ln(\sqrt{{\bf v}_1} + \sqrt{{\bf v}_1 - 1})$, $\tilde t_{1,2}=\tilde t_b\pm2\ln(\sqrt{{\bf v}_1}+\sqrt{{\bf v}_1-1})$.}, 
but in the period $\bar t\in(\bar t_1,\bar t_2)$ diverges, being the index of the $H_1^2/\dot\phi$ function in the form of a high peak ($\delta$-function): $\varepsilon \simeq (\ln\vert\dot x\vert)_{,N}$, which gives a local detail (bump) in the spectrum of $q_k$ at $k_b=H_1a(t_b)$. 

When $\mathrm c < \omega$ ($v_1>\frac17$, ${\bf v}_1>e$, $l\in(1,4. 5)$), the VA trajectory moves to the pole $x=1$ from the left ($\phi < \phi_1$), approaching it radially along $\dot\phi \simeq \frac{\mathrm m(\phi_1-\phi)}{\sqrt{{\bf v}_1}+\sqrt{{\bf v}_1-1}}$ which formally take an infinite time.

\section{Power spectra of perturbations}

At $k\ll k_0$, the spectrum has a power-law  (red) form in agreement with the observations (see (\ref{eq35})):
\begin{equation}
q_k = \bar q_0 \left(\frac{k_0}{k}\right)^{n_0},\quad 
r_k = \bar r_0 \left(\frac{k}{k_0}\right)^{2n_0},\quad 
\phi = \bar{\phi_0} \left(\frac{k}{k_0}\right)^{n_0},
\label{asi}
\end{equation}
where $\bar q_0 = \frac{\mathrm H_0}{2\pi n_0\bar\phi_0} \simeq 10^{-\!5}(\frac{k_c}{k_0})^{n_0}$, $\bar r_0 = 4n_0^2\bar\phi_0^2$, $\bar{\phi_0} = \phi_c(\frac{k_0}{k_c})^{n_0}$. The extension of the spectrum to the scales of hundreds of kiloparsecs or less depends on a value of the third constant ${\bf v}_1$, leading to an additional power in the form of a ``bump'' and/or a blue density perturbation spectrum.

At ${\bf v}_1<1$, the ratio of spectra $r_k$ grows with increasing $k$ and becomes large ($\sim1$) at $k= k_r\sim k_1$. When $k=k_p>k_1$, a high peak appears in $q_k$, potentially leading to the birth of the PBH. At higher wavenumbers $k>k_p$, the spectra fall off due to oscillations of the $\phi$ field near the side pole. At ${\bf v}_1\ll1$ the oscillations lead to instabilities: the birth of field particles with mass $\mathrm m$ and their further decay which may result in the hydrodynamical phase of the radiation-dominated plasma. This, of course, refers to the decay of the oscillatory part of the field $\varphi$, but all fields included in the vacuum polarizations remain stable (in particular, the state $V_1$ is conserved).

For ${\bf v}_1>1$ and $x\in(0,0.9)$ the perturbation spectra are two-power-law:
\begin{equation}
q_k = \frac{q_0\mathrm v^{\frac32}}{xy},\quad
r_k = r_0 \left(\frac{xy}{\mathrm v}\right)^{2},\quad
\hat x^{\frac{1}{n_0}} \hat y^{-\frac{1}{n_1}} = \frac{k\,e^{\frac{\phi^2 - \phi_{01}^2}{2}} }{k_0\sqrt{\mathrm v}},
\label{eq39}
\end{equation}
\[
n_k \simeq - \frac{n_0}{\mathrm v} \left(1 + 3x^2\left(1 - 2\frac{\mathrm v_1}{\mathrm v}\right)\right) \in \left(-n_0,n_1\right),\quad
n_1 = \frac{2n_0}{\mathrm v_1},
\]
where $q_0=\bar{q_0}(\sqrt{2e})^{\mathrm v_1-1}$, 
$r_0 = \bar{r_0}(2e)^{1-\mathrm v_1}$,
$\hat x = \frac{\phi}{\phi_{01}} =\sqrt2x$, 
$\hat y=2y$, $n_1 \in (0.04, 0.7)$. 
For $k < k_0$ ($x < 0.6$), eqs.(\ref{eq39}) correspond to the asymptotics (\ref{asi}). 
For $k > k_0 > k_1$ ($y\in(0.2,\sqrt{\frac{\mathrm v_1}{1+\mathrm v_1}})$), the spectra are blue:
\begin{equation}
q_k = q_1 \left(\frac{k}{k_1}\right)^{n_1},\quad
r_k = r_1 \left(\frac{k_1}{k}\right)^{2n_1},\quad
x^2 = 1 - \left(\frac{k_1}{k}\right)^{\!n_1\!},
\label{eq40}
\end{equation}
where $q_1=q_0\mathrm v_1^{3/2}$, $r_1 = \frac{r_0}{\mathrm v_1^2}$,  $k_1=k_0\sqrt{\mathrm v_1}(\sqrt{\frac2e})^{\phi_{01}^2}$.

A further growth of $q_k$ depends on the value of the residual vacuum. At $\vert\mathrm x \vert < \frac12$ the equation has the form 
\begin{equation}
\ddot\delta + \mathrm m\sqrt{4{\bf v}_1 + 3\delta^2} \dot\delta+\mathrm m^2\delta = 0,
\end{equation}
where $\delta=\phi-\phi_1=-\phi_1\mathrm x$.
At ${\bf v}_1\in(1,e)$, a bump appears in the spectrum (\ref{eq40}) at wavenumber $k_b>k_1$, which has the form of a single peak with its amplitude $q_b\,{}^<_\sim\, 1$\,\footnote{The peak can be approximated by a Gaussian (cf. \cite{tkachev24a}), at $k>k_b$ the spectrum has again the power-law form $q_k\sim k^{n_2}$. The bump at $k\simeq k_b$ belongs to the growing part of the three-power-law spectrum $q_k$ with the indices 
$n_k=(- n_0,n_1,n_2)$.}.
At ${\bf v}_1\in(e,e^3)$ the bump in $q_k$ is absent and the spectrum has a three-power-law form with indices $n_k=(- n_0, n_1, n_2)$, where $n_2\simeq n_1$. 
Finishing the first stage of the CVR, the field freezes at the pole $V_1$ during formally an infinite time, $\phi\rightarrow\phi_1$. After the state $V_1$ the second stage of CVR with a new field begins (and so on in the cascade of stages of the potential $V$). 

These changes in the spectrum at $k > 10$ Mpc$^{-1}$ may lead to both the early star formation (as indicated by data from the JWST) and the birth of PBHs and early SMBHs (as indicated by  LIGO and \cite{SMBH}, respectively). The extension of the spectrum to smaller scales can be restored if the observations of the JWST as well as fate of the halo mass functions and evolution of subhalos in massive galaxies will be confirmed (see \cite{tkachev24a, eroshenko24, tkachev24b} and references therein). The subsequent more accurate observational data can fix the characteristics of the power spectrum at the small scales.

\section{The entrance and exit from the vacuum attractor}
In the neighborhood of the central pole $\vert\phi\vert<\phi_0$ and $\vert\dot\phi\vert<H_0$, the solution of eq.~(\ref{eqmain}) is the sum of increasing and decreasing exponentials in time:
\begin{equation}
\phi= C_0 a^{n_0} + \bar C_0 a^{-\bar n_0},\quad 
a=\frac{k}{H_0}\propto e^{H_0t},
\end{equation}
where $\bar n_0 = 3 + n_0$. The constants $C_0$ and $\bar C_0$ are proportional to  $C$, and their ratio is in the scale when the exponential terms are equal $k_i=H_0a_i$ (beginning/boundary/entrance in the VA):
\[
\phi = \phi_i\kappa^{n_0} \left(1 \pm \vartheta^2\kappa^{-n_*}\right),\quad
n = n_0\frac{1\mp\vartheta\kappa^{-n_*}}{1\pm\vartheta^2\kappa^{-n_*}},\quad 
\varepsilon = n_0\frac{1\pm\kappa^{- n_*}}{1\mp \vartheta \kappa^{-n_*}},
\]
where $\kappa=\frac{k}{k_i}=e^{H_0(t-t_i)}$, $n_*=n_0+\bar n_0=\sqrt{3(3+4\mathrm n_0)}$, $\vartheta=\frac{n_0}{\bar n_0}\simeq0.005$. The upper sign refers to the right ($\phi_i>0$) or left ($\phi_i<0$) quadrants of the phase plane bounded by VA, and the lower sign -- to the upper or lower quadrants.  

In the vicinity of the central point, the field trajectories are under the gravitational influence of the giant swirl of the pole, which draws the trajectories to the partial automodel solution of the VA at $\kappa\sim1$. Coming to the pole from the $\vert\beta\vert\sim1$ side, the trajectories move along the VA, approaching the VA at $\kappa>1$, and turn near the center, moving away from it together with the VA towards $\vert\phi\vert\sim\phi_1/2$:
\begin{equation}
n=n_0\left(1\mp\frac{\vartheta\left(1+\vartheta\right)}{\kappa^{n_*}}+O\left(\frac{\vartheta^3}{\kappa^{2n_*}}\right)\right),\quad
\varepsilon=n_0 \left(1\pm\frac{1+\vartheta}{\kappa^{n_*}} + O\left(\frac{\vartheta}{\kappa^{2n_*}}\right)\right).
\end{equation}
In the evolutionary trajectory of the solution, only the growing exponent remains, since the decaying exponent quickly becomes insignificant. This means that the solution enters the VA near $V_0$ initially with $\phi\in(\phi_i,\phi_0)$, $\kappa\in(1,e^{N_{0i}})$ and the constant indices $n\simeq\varepsilon\simeq n_0$ (see~eqs.(28), (29)):
\begin{equation}
H=H_0\left(1-\frac{n_0\phi^2}{2}\right),\quad\phi=\phi_i\kappa^{n_0}\simeq\phi_0e^{n_0\left(N+N_0\right)},
\end{equation}
where  $N_{0i}\simeq57 \ln(\frac{\phi_0}{\phi_i})$. The power spectrum index is weakly dependent on $\phi$ ($q_k\propto H^2/\phi$, $n_k\simeq-n_0$) and agrees well with the observational model (\ref{eq14}) with $\phi_c\in(\phi_i,\phi_0)$, $k_0\simeq0.05 \exp{(N_{0c})}$ Mpc$^{-1}$, $N_{0c}\simeq57\ln(\frac{5\sqrt{1-\mathrm v_1}}{\phi_c})$. The extension of VA into the neighborhood $V_1$ depends on a single free parameter $\mathrm v_1$, which is not yet constrained by the present data.

Near the side pole, when $\vert\mathrm x\vert<0.5$, $\vert\dot\phi\vert<H_1$ and ${\bf v}_1>1$, the solution of Eq.~(\ref{eqx}) is the sum of two decaying exponentials:
\begin{equation}
\phi=C_1 a^{-\bar n_1}+C_2 a^{-\bar n_2},\quad 
a=\frac{k}{H_1}\propto e^{H_1t},
\end{equation}
where $\bar n_{1,2}=-\varepsilon_{1\pm}=\frac32(1\mp\omega)$. 
As we see, the first stage of VA leads the field to one of the side poles with $\phi=\pm\phi_1$ and $\dot\phi=0$, where the field appears in a thermal bath formed by quantum fluctuations of all fields, i.e. in a thermostat with the temperature $T=\frac{H_1}{2\pi}$, during formally an unlimited time \cite{GibbonsHawking, Starobinsky, Volovik}. This situation differs from De Sitter space-time, since the landscape of $V$ is not a constant (the state $V_1$ is stable for $\phi$, but not for all fields). Therefore, there will always be a field (not $\phi$) that will continue the CVR from $V_1$ to a lower state $V_2<V_1$. The CVR itself is the VA: it can be entered once, as in the state with $V_0$ for our trajectory of the Universe, the CVR (as well as VA) finishes when the landscape ends.   

\section{Conclusions}

The model of CRV is suggested by the observational data of recent years and can be called ``natural'' since it unties the three known parameters independent of each other (spectrum index, power spectra ratio, and number of N-epochs of evolution). 
The epoch of the growth of $\phi$ from zero is characterized by a small ratio of the power spectra ($r_k<0.01$) and a constant density perturbation spectrum index ($n_k\simeq-0.02$). 
As long as $r_c$ is unknown, the observational scales refer to any moment in time with 
$\varphi_c<0.8\,M_P$.
The OMCR is built on the current data and provides a power-law red spectrum at $k<10$ without requiring information about a potential. The theoretical VA model has two known constants, and a third (${\bf v}_1$) is necessary to extend the power spectrum at $k>10$. The comparison of VA with OMCR leads to the concept of CVR as a generator of the evolving Universe, which solves the problems of the observable cosmology.

The CRV in the expanding Universe is a sequence of evolutionary epochs of polarized vacuum density decrease determing by the dominating fields (each in its own time) going from initial zero states to non-zero values, from the dominating scalar field in the early Universe to the subsequent ones, including the $\Lambda$-term in the modern Universe. The CRV can create the entire observable cosmology -- from the Friedman model with small perturbations of the metric with non-power-law spectra, from which PBHs could have arisen, and gravitational waves in a wide wavenumber range, to the formation of dark matter and dark energy, early galaxies and SMBHs, the large-scale structure of the Universe. 

\begin{acknowledgement}
Authors cordially thanks P.B. Ivanov for the assistance in preparing the paper. 
\end{acknowledgement}

\end{document}